\documentclass[sigconf]{acmart}

\usepackage{multirow}
\newsavebox{\gplanpanelbody}
\AtBeginDocument{%
  }

\usepackage{cprotect}
\ifdefined\cprotect\else\expandafter\newcommand\csname cprotect\endcsname[1][]{}\fi

\copyrightyear{2026}
\acmYear{2026}
\setcopyright{cc}
\setcctype{by-nc-nd}
\acmConference[RecSys '26]{20th ACM Conference on Recommender Systems}{September 27--October 02, 2026}{Minneapolis, MN, USA}
\acmBooktitle{20th ACM Conference on Recommender Systems (RecSys '26), September 27--October 02, 2026, Minneapolis, MN, USA}
\acmDOI{10.1145/3773078.3831861}
\acmISBN{979-8-4007-2284-4/26/09}

\usepackage{acmart-taps}
\usepackage{tabularx}
\usepackage{colortbl}
\usepackage{array}
\usepackage{xcolor}

\usepackage{gplan-taps-layout}

\begin{document}

\title{Generative Spatiotemporal Intent Sequence Recommendation via Implicit Reasoning in Amap}


\author{Sicong Wang}
\authornote{Contributed equally to this research.}
\affiliation{%
  \institution{AMAP, Alibaba Group}
  \city{Beijing}
  \country{China}
}
\email{wsc488938@alibaba-inc.com}
\orcid{0009-0000-0757-278X}

\author{Ruiting Dong}
\authornotemark[1]
\affiliation{
  \institution{AMAP, Alibaba Group}
  \city{Beijing}
  \country{China}
}
\email{dongruiting.drt@alibaba-inc.com}
\orcid{0009-0003-4913-8297}

\author{Yue Liu}
\authornotemark[1]
\affiliation{%
  \institution{AMAP, Alibaba Group}
  \city{Beijing}
  \country{China}
}
\email{ly355576@alibaba-inc.com}
\orcid{0009-0008-3977-0148}

\author{Bowen Zheng}
\authornotemark[1]
\affiliation{
  \institution{AMAP, Alibaba Group}
  \city{Beijing}
  \country{China}
}
\email{zbw405521@alibaba-inc.com}
\orcid{0009-0003-3044-1472}

\author{Jun Meng}
\affiliation{%
  \institution{AMAP, Alibaba Group}
  \city{Beijing}
  \country{China}
}
\email{xiuxian.mj@alibaba-inc.com}
\orcid{0009-0000-3852-6380}

\author{Jie Li}
\affiliation{%
  \institution{AMAP, Alibaba Group}
  \city{Beijing}
  \country{China}
}
\email{lj313796@alibaba-inc.com}
\orcid{0009-0008-7138-700X}

\author{Shuaijun Guo}
\affiliation{
  \institution{AMAP, Alibaba Group}
  \city{Beijing}
  \country{China}
}
\email{guoshuaijun.gsj@alibaba-inc.com}
\orcid{0009-0008-9213-4917}

\author{Yu Gu}
\affiliation{%
  \institution{AMAP, Alibaba Group}
  \city{Beijing}
  \country{China}
}
\email{gy507704@alibaba-inc.com}
\orcid{0009-0004-1055-6093}

\author{Fanyi Di}
\authornote{Corresponding author.}
\affiliation{%
  \institution{AMAP, Alibaba Group}
  \city{Beijing}
  \country{China}
}
\email{difanyi.dfy@alibaba-inc.com}
\orcid{0009-0009-5272-7934}

\author{Xin Li}
\authornotemark[2]
\affiliation{%
  \institution{AMAP, Alibaba Group}
  \city{Beijing}
  \country{China}
}
\email{beilai.bl@alibaba-inc.com}
\orcid{0000-0001-6888-7064}
\renewcommand{\shortauthors}{Wang et al.}

\begin{abstract}
Real-world user behavior rarely consists of isolated actions; instead, it often forms intent flows governed by spatiotemporal dependencies. To provide integrated service recommendations, we focus on the task of \textbf{G}enera\-tive \textbf{S}patio\-temporal \textbf{I}ntent \textbf{S}equence \textbf{R}ecom\-mendation (\textbf{GSISR}), which aims to generate intent sequences that are logically coherent and physically executable within complex spatiotemporal contexts. While LLMs offer strong reasoning potential for GSISR, direct industrial deployment is limited by high inference latency and context-mismatched or physically infeasible plans. To address these challenges, we propose a generative framework, GPlan, that internalizes LLM reasoning into lightweight models through two components. First, to enable reasoning under strict latency constraints, we introduce Progressive Implicit CoT Distillation, which compresses explicit reasoning processes into reserved latent tokens, allowing small models to inherit complex planning logic without generating long reasoning text. Second, to address the disconnect between general knowledge and real-world constraints, we design Spatiotemporal Counterfactual DPO. By aligning the model with counterfactual context-plan pairs, we improve sensitivity to spatiotemporal context and reduce context-mismatched plans. Offline experiments and online A/B testing demonstrate that our approach improves sequence coherence and context responsiveness. Our implementation and the anonymized GSISR dataset are available at \url{https://github.com/alibaba/GPlan}.
\end{abstract}



\maketitle

\section{Introduction}
Modern mobile map applications such as Amap\footnote{\url{https://amap.com/}} bridge the digital and physical worlds. In Amap homepage recommendation, the system employs an agent to recommend various tool cards, including ride-hailing, destination exploration, and local lifestyle services. The agent identifies potential intents from user profiles, historical interactions, and spatiotemporal contexts, and then invokes specific tools that return POIs or content. These results are ultimately presented as a sequence of cards. Throughout this paper, an \emph{intent} denotes a parameterized invocation of an Amap homepage tool (e.g., ride-hailing with origin and destination arguments), and a \emph{card} denotes the resulting UI surface; the output of GPlan is therefore an ordered sequence of intents, which the production stack renders as cards.

Real-world user behaviors are rarely isolated actions; they manifest as intent flows governed by spatiotemporal dependencies \cite{10.1145/3639048}, e.g., a user in cross-city travel may follow: \textit{[Request a ride]$\rightarrow$[Hotel check-in]$\rightarrow$[Nearby dining]$\rightarrow$[Return to hotel]}. To provide integrated service recommendations, we define \textbf{Generative Spatiotemporal Intent Sequence Recommendation (GSISR)}: generating intent sequences that are both logically coherent and physically executable under spatiotemporal constraints.

Existing sequential recommenders primarily optimize next-item prediction~\cite{10.1007/s11704-025-41329-w,BOKA2024102427}, which does not ensure the global coherence required by GSISR and can lead to logical inconsistencies or repetitive loops in long intent sequences.
The rise of LLMs has catalyzed a shift toward generative recommendation \cite{wu2024survey,10.1145/3678004}, yet direct LLM deployment faces two key challenges: (1) \textit{Computational bottleneck}---Chain-of-Thought (CoT) \cite{10.5555/3600270.3602070} reasoning incurs latency that exceeds production requirements; (2) \textit{Spatiotemporal hallucination}---LLMs may lack access to application-specific spatiotemporal constraints, producing physically infeasible itineraries.

\begin{figure*}[t]
  \centering
  \includegraphics[width=\linewidth]{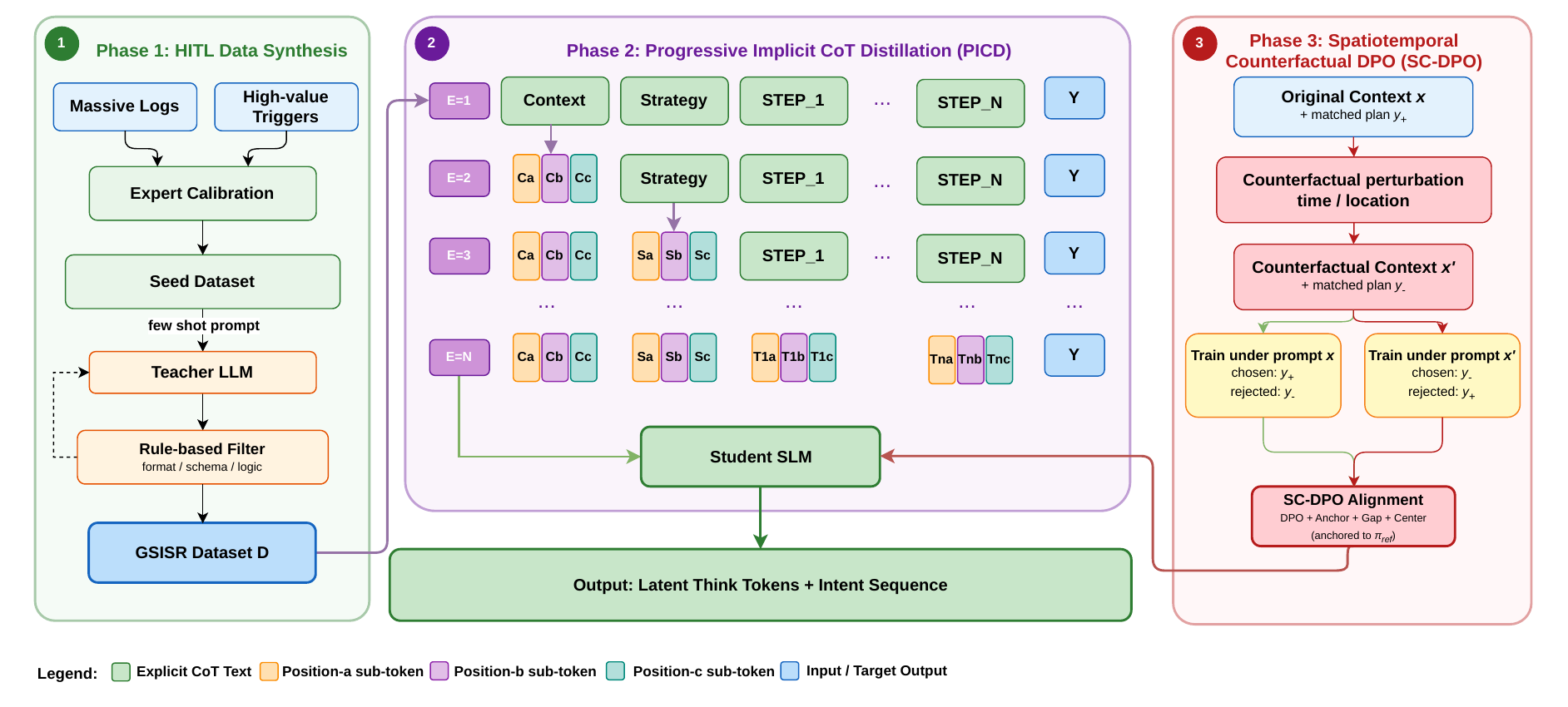}
  \Description{A left-to-right, three-phase diagram of GPlan. Phase 1
  constructs the GSISR dataset from expert-calibrated seed examples, teacher
  generation, and rule-based filtering. Phase 2 progressively replaces the
  context, strategy, and step-level chain-of-thought blocks with
  position-specific latent-token triplets to train a student SLM, which
  outputs latent think tokens followed by an intent sequence. Phase 3
  perturbs time or location to form matched and counterfactual context-plan
  pairs and aligns the student with a reference-anchored SC-DPO objective.}
  \caption{The three-phase GPlan pipeline. (1) HITL data synthesis uses expert-calibrated seeds, a teacher LLM, and rule-based filtering to construct the GSISR dataset. (2) PICD progressively replaces context-, strategy-, and step-level CoT blocks with position-specific latent-token triplets, yielding a student that emits latent think tokens followed by an intent sequence. (3) SC-DPO perturbs time or location to create bidirectional matched/mismatched context--plan preferences and aligns the student with the reference-anchored objective.}
  \label{fig:framework}
\end{figure*}

To address these challenges, we propose \textbf{GPlan} (``Guess What You Plan''), a framework that internalizes LLM-level planning into lightweight industrial models. GPlan comprises two core components: (1) \textbf{Progressive Implicit CoT Distillation (PICD)}, which progressively compresses explicit teacher reasoning into reserved latent tokens, preserving expert-level logic while meeting online latency requirements; (2) \textbf{Spatiotemporal Counterfactual DPO (SC-DPO)}, which constructs bidirectional preference pairs between matched and counterfactually-perturbed spatiotemporal contexts and applies a reference-anchored DPO objective \cite{rafailov2023direct} so that the model learns to change its plan when the context shifts. The framework is illustrated in Figure~\ref{fig:framework}.

The main contributions are:
\begin{itemize}
\item We define the GSISR task and release an anonymized industrial dataset from Amap for research on logical coherence and physical feasibility in recommendation.
\item We design PICD, a latent CoT distillation built around a multi-token reasoning scaffold and a \emph{compression-aware learning rate} (CALR) that ties LR transitions to the state of the latent compression curriculum, enabling efficient student inference while preserving the implicit-CoT interface.
\item We propose SC-DPO, a counterfactual preference-alignment objective whose reference-anchor, bounded-margin, and centered-drift terms introduce context-conditional preference without degrading the SFT-acquired planning ability.
\end{itemize}

\section{Related Work}\label{sec:rw}

\paragraph{Sequential, location, and generative recommendation.} Sequential recommenders model ordered behavior to predict future interactions. SASRec and BERT4Rec use Transformer encoders to capture short- and long-range dependencies~\cite{SASRec,BERT4Rec}, while ReaRec explores latent multi-step computation~\cite{ReaRec}. In LBS, next-location and next-POI models incorporate spatiotemporal relations and trajectory transitions to predict the next destination~\cite{luo2021stan,yang2022getnext}, while itinerary recommenders construct ordered POI routes under user-interest and travel-time constraints~\cite{lim2017personalized}. Generative recommendation reformulates outputs as natural-language tokens or semantic identifiers: P5~\cite{P5} unifies recommendation tasks under a text-to-text paradigm, TIGER~\cite{TIGER} applies generative retrieval via autoregressively decoded semantic IDs, and HSTU~\cite{HSTU} scales generative transducers to industrial settings. Surveys further summarize how LLMs support semantic understanding, instruction following, and generative retrieval in recommender systems~\cite{wu2024survey,10.1145/3678004,10.1007/s11704-025-41329-w,BOKA2024102427}. Together, these lines center on next-item retrieval, next-place prediction, or POI-only routes. GSISR instead generates a heterogeneous, parameterized sequence of tool invocations in which later actions may depend on earlier arguments and the current spatiotemporal context.

\paragraph{Efficient reasoning and preference alignment.} Chain-of-thought prompting improves LLM reasoning by surfacing intermediate steps before the final answer~\cite{10.5555/3600270.3602070}, but explicit CoT decoding is too slow for production. One line of work distills reasoning into smaller students that still emit explicit traces~\cite{magister2023teachingsmalllanguagemodels,li2024mixeddistillationhelpssmaller,li2025smallmodelsstrugglelearn,luo2025deconstructinglongchainofthoughtstructured,liao2024textitskinterninternalizingsymbolicknowledge,feng2025efficient}; this preserves the reasoning text but not its decoding cost. A more recent line moves reasoning into latent or compressed representations: ICoT-KD~\cite{deng2023implicitchainthoughtreasoning} distills an explicit-CoT teacher into layer-wise hidden-state computation; Stepwise Internalization~\cite{deng2024explicitcotimplicitcot} gradually removes intermediate reasoning steps; CCoT~\cite{cheng2024compressedchainthoughtefficient} generates variable-length continuous contemplation tokens; CODI~\cite{shen-etal-2025-codi} self-distills explicit CoT into a continuous representation; and Coconut~\cite{hao2025training} feeds the last hidden state back as the next input embedding. PICD builds on this line with typed, position-specific token triplets, a plan-length-dependent forward compression curriculum, and CALR for variable-length intent-sequence generation. For preference alignment, DPO~\cite{rafailov2023direct} learns from preference pairs without a separate reward model, while S-DPO adapts this objective to recommendation ranking with multiple negatives~\cite{chen2024softmax}. Counterfactual sequential recommenders synthesize alternative interaction sequences to improve data coverage or representation robustness~\cite{wang2021counterfactual,zhang2021causerec}. In LBS recommendation, however, a fluent and semantically plausible plan may still violate temporal, spatial, or service-state constraints. SC-DPO therefore constructs bidirectional preference reversals for minimally changed time or location contexts and constrains their reward shifts relative to the SFT policy.

\section{Preliminary}\label{sec:prelim}
\subsection{Problem Formulation}

The GSISR task aims to generate an executable intent sequence that satisfies complex user needs in the Amap homepage agent.

\textbf{Input Space:} Let $\mathcal{E}$ denote the intent library. An input instance is $x=(u,\mathcal{H},s)\in\mathcal{X}$, where $u$ is the user profile, $\mathcal{H}$ the interaction history, and $s$ the spatiotemporal context.

\textbf{Output Space:} The goal is to generate an ordered sequence of $n$ parameterized intents $\mathbf{Y}=(a_1,a_2,\dots,a_n)\in\mathcal{Y}$. Each intent $a_k=(v_k,p_k)$ invokes a tool $v_k\in\mathcal{E}$ with arguments $p_k$.

\textbf{Objective:} Our goal is to learn a generative model $f_\theta:\mathcal{X}\rightarrow\mathcal{Y}$. Formally, we aim to find the optimal parameters $\theta^*$ that maximize the log-likelihood of generating curated target sequences from the dataset $\mathcal{D}$:
\begin{equation}
    \theta^* = \arg\max_\theta \sum_{(x,\mathbf{Y})\in\mathcal{D}} \sum_{\ell=1}^{L} \log P(y_\ell \mid y_{<\ell},x;\theta)
\end{equation}
where $L$ is the length of the serialized sequence $\mathbf{Y}$ and $y_\ell$ denotes its $\ell$-th token. Crucially, the dataset $\mathcal{D}$ is constructed (see Sec.~\ref{sec:data_synthesis}) to ensure that each curated target sequence $\mathbf{Y}$ satisfies the physical constraints imposed by the spatiotemporal context $s$ and the logical dependencies encoded by $\mathcal{E}$.

\aptLtoX{\par\smallskip
\noindent
\begingroup
\small
\setlength{\tabcolsep}{6pt}
\renewcommand{\arraystretch}{1.1}
\begin{tabular}{|>{\raggedright\arraybackslash}l{\dimexpr\columnwidth-14pt\relax}|}
  \hline
  \rowcolor{gray!65!black}\color{white}\bfseries Example: Cross-City Travel Anchor \\
  \hline
  \rowcolor{gray!4!white}
  \textbf{Context} $x$: a cross-city traveler has booked a 
  hotel in city B and just landed at city B airport at 15:00. \\
  \rowcolor{gray!4!white}
  \textbf{GPlan output} $\mathbf{Y}$:\\
  \begin{enumerate}\itemsep1pt
  \item Ride-hail to the hotel
  \item Hotel check-in reminder
  \item Food near the hotel
  \item Leisure venues near the hotel
  \item City-wide scenic spots
  \end{enumerate} \\  
  \rowcolor{gray!4!white}
  \textbf{Listwise baseline output:}\\
  \begin{enumerate}\itemsep1pt
  \item Ride-hail to the hotel
  \item Hotel check-in reminder
  \item Food / scenic spots nearby \emph{(still at the airport)}
  \item Nearby public transit
  \item Today's weather
  \end{enumerate} \\  
  \rowcolor{gray!4!white}
  Once the anchor intent (heading to the hotel) is fixed, GPlan re-bases
  downstream POIs onto the hotel area where the user will actually
  arrive, whereas a listwise baseline keeps scoring candidates relative
  to the current location (the airport) and falls back to generic
  support cards. \\
  \hline
\end{tabular}
\endgroup
\par\smallskip}{\par\smallskip
\begingroup
\setlength{\fboxsep}{6pt}
\setlength{\fboxrule}{0.5pt}
\begin{lrbox}{\gplanpanelbody}
\begin{minipage}{\dimexpr\columnwidth-2\fboxsep-2\fboxrule-12pt\relax}
\small
\textbf{Context} $x$: a cross-city traveler has booked a
hotel in city B and just landed at city B airport at 15:00.

\textbf{GPlan output} $\mathbf{Y}$:
\begin{enumerate}\itemsep1pt
\item Ride-hail to the hotel
\item Hotel check-in reminder
\item Food near the hotel
\item Leisure venues near the hotel
\item City-wide scenic spots
\end{enumerate}

\textbf{Listwise baseline output:}
\begin{enumerate}\itemsep1pt
\item Ride-hail to the hotel
\item Hotel check-in reminder
\item Food / scenic spots nearby \emph{(still at the airport)}
\item Nearby public transit
\item Today's weather
\end{enumerate}

Once the anchor intent (heading to the hotel) is fixed, GPlan re-bases
downstream POIs onto the hotel area where the user will actually
arrive, whereas a listwise baseline keeps scoring candidates relative
to the current location (the airport) and falls back to generic
support cards.
\end{minipage}
\end{lrbox}

\noindent
\begin{minipage}{\columnwidth}
\fcolorbox{gray!65!black}{gray!65!black}{%
  \makebox[\dimexpr\columnwidth-2\fboxsep-2\fboxrule\relax][l]{%
    \hspace*{6pt}\color{white}\bfseries\small
    Example: Cross-City Travel Anchor}}%
\par\nointerlineskip
\kern-\fboxrule
\noindent
\fcolorbox{gray!65!black}{gray!4!white}{%
  \makebox[\dimexpr\columnwidth-2\fboxsep-2\fboxrule\relax][c]{%
    \usebox{\gplanpanelbody}}}%
\end{minipage}
\endgroup
\par\smallskip}

\subsection{Human-in-the-loop Data Synthesis}\label{sec:data_synthesis}
No single map app observes all of a user's activity, and the actions it records need not form a coherent plan. Real-world LBS logs are thus sparse and noisy, making them impractical as direct training data for GSISR. We therefore design a \textbf{Human-in-the-loop (HITL)} pipeline to construct the GSISR dataset with logical depth and physical grounding.

We first extract \textbf{high-value triggers} from large-scale logs --- log patterns that imply multi-step service needs, such as cross-city arrivals or active ride/hotel orders. After cleaning, domain experts at Amap perform logical calibration on the resulting trajectories, yielding a \textbf{Seed Dataset} of 1{,}000 ``Intent--Environment--Logic'' triplets that anchor the downstream synthesis with human-validated examples.

To scale beyond the seed, we employ a teacher model (Qwen3-235B-A22B~\cite{yang2025qwen3technicalreport}) with few-shot prompting~\cite{brown2020languagemodelsfewshotlearners} and the Seed Dataset as in-context examples. The teacher autoregressively generates a structured reasoning path $\mathbf{R}$ and intent sequence $\mathbf{Y}$ conditioned on user profile $u$ and spatiotemporal context $s$.

Each generated sample then passes through a \textbf{rule-based filter} and is removed if any tier fails: (i) \emph{format} checks --- JSON parsability and a valid plan length; (ii) \emph{schema} checks --- every intent must use a tool from a fixed library, with required parameters present and enumerable fields restricted to their admissible values; and (iii) \emph{logic} invariants --- requiring every parameter reference (e.g., a destination consumed by a downstream action) to resolve to a producer intent earlier in the sequence, and bounding the count of mutually exclusive travel-class intents (e.g., at most one ride-hailing call per sequence). Across our synthesized candidates, $28.9\%$ are removed, leaving $100{,}000$ training samples and $1{,}000$ test samples in the final dataset $\mathcal{D}$. The same filter is re-applied at evaluation time as a minimum validity check on any model's output (Sec.~\ref{sec:experiments}).

\paragraph{Label Provenance.} The 1{,}000 seed triplets are calibrated by Amap domain experts. The remaining targets are generated by Qwen3-235B-A22B from these examples and filtered by the format, schema, and logic checks above. The expert-calibrated seeds and deterministic filter are independent of the student model, although most training and offline evaluation targets are teacher-generated.

\section{Methodology}
GPlan combines two components for low-latency, context-sensitive intent-sequence generation. PICD compresses explicit teacher reasoning into latent tokens, while SC-DPO trains the model to revise its plan when the time or location changes.

\subsection{Progressive Implicit CoT Distillation}\label{sec:picd}

PICD progressively compresses structured teacher CoT into reserved latent tokens for low-latency, variable-length intent-sequence generation. This process combines three mechanisms: (i) a typed multi-token scaffold that preserves the semantic organization of the teacher's CoT, (ii) a forward block-wise compression schedule whose block count follows the plan length, and (iii) a compression-aware learning-rate schedule that aligns LR transitions with the changing reasoning representation. A section-normalized loss keeps JSON-plan supervision at a fixed section-level weight as the textual CoT prefix shrinks.

\subsubsection{Expert Demonstration with Structured Reasoning}

We synthesize an expert dataset in which the teacher LLM produces a structured reasoning path $\mathbf{R}$ before the final intent sequence. $\mathbf{R}$ has three \emph{types} of semantic blocks: \texttt{<CONTEXT>} (contextual analysis), \texttt{<STRATEGY>} (planning strategy), and \texttt{<STEP\_$i$>} (per-intent justification, repeated once per item in the JSON plan).

\cprotect[mm]\aptLtoX{\par\smallskip
\noindent
\begingroup
\small
\setlength{\tabcolsep}{6pt}
\renewcommand{\arraystretch}{1.1}

\begin{tabular}{|>{\raggedright\arraybackslash}l{\dimexpr\columnwidth-14pt\relax}|}
  \hline
  \rowcolor{gray!75!black}\color{white}\bfseries 
  Chain-of-Thought Structure \\
  \hline
  \rowcolor{gray!5!white}
  \ttfamily
  <THOUGHT> \\
  \rowcolor{gray!5!white}
  \ttfamily <CONTEXT>Briefly analyze the current context and \\
  \rowcolor{gray!5!white}
  \ttfamily the user's potential needs</CONTEXT> \\
  \rowcolor{gray!5!white}
  \ttfamily <STRATEGY>Based on the context analysis, devise \\
  \rowcolor{gray!5!white}
  \ttfamily the core strategy for the plan</STRATEGY> \\
  \rowcolor{gray!5!white}
  \ttfamily <STEP\_1>Focus on analyzing the primary and most \\
  \rowcolor{gray!5!white}
  \ttfamily crucial intent</STEP\_1> \\
  \rowcolor{gray!5!white}
  \ttfamily ... \\
  \rowcolor{gray!5!white}
  \ttfamily <STEP\_n>Explain why the n-th intent is \\
  \rowcolor{gray!5!white}
  \ttfamily recommended</STEP\_n> \\
  \rowcolor{gray!5!white}
  \ttfamily </THOUGHT> \\
  \hline
\end{tabular}
\endgroup
\par\smallskip}{\par\smallskip
\begingroup
\setlength{\fboxsep}{6pt}
\setlength{\fboxrule}{0.5pt}
\begin{lrbox}{\gplanpanelbody}
\begin{minipage}{\dimexpr\columnwidth-2\fboxsep-2\fboxrule-12pt\relax}
\begin{verbatim}
<THOUGHT>
<CONTEXT>Briefly analyze the current context and
the user's potential needs</CONTEXT>
<STRATEGY>Based on the context analysis, devise
the core strategy for the plan</STRATEGY>
<STEP_1>Focus on analyzing the primary and most
crucial intent</STEP_1>
...
<STEP_n>Explain why the n-th intent is
recommended</STEP_n>
</THOUGHT>
\end{verbatim}
\end{minipage}
\end{lrbox}

\noindent
\begin{minipage}{\columnwidth}
\fcolorbox{gray!75!black}{gray!75!black}{%
  \makebox[\dimexpr\columnwidth-2\fboxsep-2\fboxrule\relax][l]{%
    \hspace*{6pt}\color{white}\bfseries\small
    Chain-of-Thought Structure}}%
\par\nointerlineskip
\kern-\fboxrule
\noindent
\fcolorbox{gray!75!black}{gray!5!white}{%
  \makebox[\dimexpr\columnwidth-2\fboxsep-2\fboxrule\relax][c]{%
    \usebox{\gplanpanelbody}}}%
\end{minipage}
\endgroup
\par\smallskip}

While structured CoT improves decision quality, its autoregressive generation cost is prohibitive under production latency constraints.

\subsubsection{Latent Multi-Token Reasoning Scaffold}\label{sec:picd_scaffold}

To reduce the autoregressive generation overhead of explicit CoT while preserving its structural information, PICD replaces each complete semantic block of $\mathbf{R}$ with a fixed-length sequence of $K$ reserved special tokens, keeping only the outer \texttt{<THOUGHT>...</THOUGHT>} wrapper. With $K=3$, the fully latent prefix is compactly written as
\begin{equation}
\Phi(\mathbf{R}) =
\texttt{<THOUGHT>}\; C\,S\,T_1\cdots T_{|\mathbf{Y}|}\;\texttt{</THOUGHT>},
\end{equation}
where $C=(c_a,c_b,c_c)$ encodes the context block, $S=(s_a,s_b,s_c)$ encodes the strategy block, and $T_i=(t_{i,a},t_{i,b},t_{i,c})$ encodes the $i$-th step block. These symbols denote distinct reserved vocabulary tokens, e.g., $c_\cdot$ corresponds to \texttt{<THOUGHT\_CONTEXT\_*>}, $s_\cdot$ to \texttt{<THOUGHT\_STRATEGY\_*>}, and $t_{i,\cdot}$ to \texttt{<T\_$i$\_*>}. Thus the inner XML tags such as \texttt{<CONTEXT>} and \texttt{<STEP\_$i$>} are removed once their blocks are compressed.

We choose $K=3$ to give each latent block enough capacity for the multi-faceted content of one reasoning step: in a \texttt{<STEP\_$i$>} block, the latent triplet jointly encodes tool selection, parameter assignment, and the link to the previous intent in the plan; in \texttt{<CONTEXT>} and \texttt{<STRATEGY>} it accommodates the user's spatiotemporal situation and the high-level plan rationale. $K$ is treated as a fixed architectural choice rather than a per-run hyperparameter.

Because the JSON plan length $|\mathbf{Y}|$ varies across samples, so does the total number of latent blocks:
\begin{equation}
B = 2 + |\mathbf{Y}|,
\end{equation}
where the constant $2$ accounts for \texttt{<CONTEXT>} and \texttt{<STRATEGY>}. We track $B$ dynamically per sample during training rather than padding to a fixed length, so samples with shorter plans require fewer blocks to reach full latent compression.

Based on the sequence lengths in our training data and practical requirements for serving efficiency and user experience, we set the maximum intent-sequence length to nine intents for training, evaluation, and online serving.

\subsubsection{Forward Progressive Compression and Compression-Aware LR}\label{sec:picd_calr}

\paragraph{Forward progressive schedule.} Across the $N$ training epochs, blocks are replaced with their latent token triplets in \emph{forward} order: \texttt{<CONTEXT>}, then \texttt{<STRATEGY>}, then \texttt{<STEP\_1>}, \texttt{<STEP\_2>}, and so on. This order matches the natural construction of an intent plan---first situate the user, then commit to a strategy, then expand into concrete steps. Let $b(e)$ denote the number of leading blocks compressed at the start of epoch $e$. Training proceeds through (i) an \emph{initial phase} ($b{=}0$) replicating full-text CoT, (ii) a \emph{progressive phase} ($0{<}b{<}B$) replacing the leading $b$ blocks with latent triplets while the rest remain in text, and (iii) a \emph{fully implicit phase} ($b{\geq}B$) with the entire reasoning prefix in latent form. Because $B$ varies per sample, samples with shorter plans enter the fully-implicit phase earlier than longer ones.

\paragraph{Non-stationary distillation.} A subtle issue arises in scheduling the optimizer through this curriculum: the prediction target is non-stationary. As $b$ grows, more of the CoT prefix becomes latent tokens, and the student must learn how those tokens encode the missing semantics. A standard monotonically decaying schedule (e.g., cosine over all $N$ epochs) shrinks the LR before that latent structure is fully acquired, leaving artifacts such as repeated triplets, missing tokens, and out-of-order sequences in the emitted reasoning. Holding the LR constant throughout stabilizes the latent structure but leaves no low-LR phase in which to refine the JSON plan against its now-final latent input.

\paragraph{Compression-Aware LR (CALR)} CALR ties the LR schedule to the state of the compression curriculum. Let $t$ index optimization steps and let $g(t)$ denote the number of leading blocks scheduled for compression at step $t$, i.e., the step-level form of the epoch schedule $b(e)$. For a sample with $B = 2 + |\mathbf{Y}|$ reasoning blocks, the prefix is fully latent once $g(t)$ reaches $B$; before that point, the model is still learning what each latent token should encode. We set the LR-transition target to $B^\star = 9$, the empirical $p_{90}$ of $B$ across our training set, so that by the time the LR drops, most samples have already seen a fully latent prefix at least once. Concretely, CALR keeps a higher LR while the latent prefix is being formed and switches to a lower cosine-decayed LR once the latent representation is essentially stable:
\begin{equation}
\eta(t) = \begin{cases}
\eta_{\text{struct}} & g(t) \leq B^\star \quad \textit{(structure acquisition)} \\
\eta_{\text{polish}} \cdot \mathrm{cosine}(t-t^\star) & g(t) > B^\star \quad \textit{(semantic polish)}
\end{cases}
\end{equation}
where $t^\star$ is the first refinement step and $\eta_{\text{struct}}>\eta_{\text{polish}}$. In our main runs $\eta_{\text{struct}}=5\!\times\!10^{-6}$ and $\eta_{\text{polish}}=1\!\times\!10^{-6}$. Intuitively, the first phase builds the latent token structure under a stable high LR; the second phase refines the JSON plan conditioned on the now-final latent prefix. Sec.~\ref{sec:abl_calr} validates this design empirically.

\paragraph{Section-normalized training loss.} PICD changes the number of supervised CoT tokens over the curriculum: early examples contain full textual reasoning, whereas later examples replace more blocks with three-token latent triplets. Under standard token-level cross entropy, the relative weight of the JSON plan would therefore vary with the compression stage. We stabilize the objective by averaging within the CoT and JSON sections separately and combining them equally:
\begin{equation}
\mathcal{L} = \frac{1}{2}\left(\mathcal{L}_{\textsc{cot}} + \mathcal{L}_{\textsc{json}}\right),
\end{equation}
where $\mathcal{L}_{\textsc{cot}}$ and $\mathcal{L}_{\textsc{json}}$ are mean token losses over their respective active sections. This keeps the JSON plan at a fixed section-level weight throughout progressive compression.

\subsection{Spatiotemporal Counterfactual DPO}\label{sec:scdpo}

After supervised fine-tuning, the model has acquired basic user understanding, trajectory interpretation, and plan generation. Static supervision alone, however, is a weak signal for the \emph{conditional} dependence of the plan on spatiotemporal context: the same user under weekday morning rush, weekend night, or a relocated current position should receive substantively different intent sequences, yet the SFT objective does not enforce this dependence. We therefore introduce \textbf{Spatiotemporal Counterfactual DPO (SC-DPO)}, a post-training step that teaches the model how a change in context should reshape the generated plan.

\subsubsection{Counterfactual Pair Construction}
For each anchor sample in the training split, we construct two related contexts: an \emph{original} context $x$ and a \emph{counterfactual} context $x'$. The two share the user profile, long-term behavior summary, and main history; they differ only in spatiotemporal variables --- current time, current location, current city, weekday/weekend flag, or holiday flag. Each context is paired with a teacher-generated, filter-passed plan produced by the pipeline in Sec.~\ref{sec:data_synthesis}. We materialize the pair bidirectionally: under prompt $x$, the $x$-matched plan is chosen and the $x'$-matched plan is rejected; under prompt $x'$, the comparison is reversed. This symmetry is essential because the rejected plan is not globally invalid; it is only wrong for the current context, and forcing the model to flip its preference when the context flips trains it to learn ``match the context'' rather than ``always avoid the counterfactual plan''.

For input $x$, the matching response is denoted $y_+$ and the response matched to the alternative context $x'$ is denoted $y_-$. Crucially, $y_-$ is a \emph{context-mismatched} response, not a globally low-quality one: it would be a perfectly valid plan under $x'$, but does not fit $x$. This distinction motivates the loss design below.

\subsubsection{Reference-Anchored Optimization Objective}

\paragraph{Why standard DPO is insufficient.} Standard DPO treats a $(y_w, y_l)$ pair as an absolute preference and rewards the policy for continually widening the log-likelihood gap. Applied directly to counterfactual pairs, this would push $\pi_\theta(y_-\mid x)$ arbitrarily low even though $y_-$ is a valid plan in its own context. The model would not only learn context-conditional preference but also degrade the general planning ability acquired during SFT.

\paragraph{Objective intuition.}
A counterfactual negative is valid in its own context, so SC-DPO must learn a local preference without treating it as globally undesirable. Under context $x$, the DPO term ranks the $x$-matched plan above the $x'$-matched plan. The anchor prevents this update from lowering the matched plan relative to the SFT policy. The gap and center controls prevent the preference from being learned by globally suppressing $y_-$: the gap keeps the margin positive but bounded, while the center keeps the pair's average reward shift near the reference policy. We now formalize these components.

\paragraph{Reward shifts.} Let $\pi_{\text{ref}}$ be the SFT model and $\pi_\theta$ the model being optimized. For each counterfactual pair $(x, y_+, y_-)$, define the reward shift of each response relative to the reference (with DPO temperature $\beta_{\text{DPO}}$):
\begin{align}
r_+ &= \beta_{\text{DPO}} \bigl[\log \pi_\theta(y_+ \mid x) - \log \pi_{\text{ref}}(y_+ \mid x)\bigr], \\
r_- &= \beta_{\text{DPO}} \bigl[\log \pi_\theta(y_- \mid x) - \log \pi_{\text{ref}}(y_- \mid x)\bigr].
\end{align}

\paragraph{Preference term.} As in DPO, the preference term encourages a positive margin between matched and mismatched responses:
\begin{equation}
\mathcal{L}_{\text{DPO}} = -\log \sigma(r_+ - r_-).
\end{equation}

\paragraph{Reference anchor.} To prevent SFT-acquired planning ability from drifting during DPO, we anchor the matched response to the reference policy with a hinge penalty (with margin $\delta$):
\begin{equation}
\mathcal{L}_{\text{anchor}} = \max(0, \delta - r_+)^2.
\end{equation}
This term requires that $\pi_\theta(y_+ \mid x)$ does not fall meaningfully below $\pi_{\text{ref}}(y_+ \mid x)$; we set $\delta=0$ in the main runs. The model is therefore not allowed to sacrifice the correct response under $x$ to amplify the margin against $y_-$.

\paragraph{Bounded margin and centered drift.} To prevent the model from gaming the preference term by collapsing $y_-$, we constrain both the preference margin $\text{gap} = r_+ - r_-$ and the pair's overall reward drift $\text{center} = \tfrac{1}{2}(r_+ + r_-)$:
\begin{align}
\mathcal{L}_{\text{gap},\ell} &= \max(0, \gamma_\ell - \text{gap})^2, \\
\mathcal{L}_{\text{gap},h} &= \max(0, \text{gap} - \gamma_h)^2,
\end{align}
\begin{equation}
\mathcal{L}_{\text{center}} = (\text{center} - m)^2.
\end{equation}
$[\gamma_\ell, \gamma_h]$ defines the desired range of the preference margin --- positive, but not unbounded --- and the two gap components allow asymmetric penalties for falling below the lower bound versus exceeding the upper bound. The center target $m$ (set to $0$ in our experiments) keeps the pair-level reward drift near the reference, so the loss does not encode a global ``$y_-$ is bad'' signal.

\paragraph{Final objective.}
\begin{equation}
\mathcal{L}_{\text{SC-DPO}} = \mathcal{L}_{\text{DPO}} + \lambda_a\,\mathcal{L}_{\text{anchor}} + \lambda_{g,\ell}\,\mathcal{L}_{\text{gap},\ell} + \lambda_{g,h}\,\mathcal{L}_{\text{gap},h} + \lambda_c\,\mathcal{L}_{\text{center}}.
\end{equation}
These loss components address distinct concerns of the counterfactual setting: $\mathcal{L}_{\text{DPO}}$ provides the preference signal; $\mathcal{L}_{\text{anchor}}$ protects the SFT-acquired planning ability on matched responses; and the gap and center components encode that the desired behavior is a bounded, context-conditional preference rather than absolute aversion to $y_-$.

Conceptually, SC-DPO teaches the model not only ``which plan is better'' but ``why the same user warrants a different plan when time or location shifts''. It is best understood as a post-training step that makes the model change its plan when the user's time, location, or trip state changes; its contribution is most pronounced on metrics sensitive to context responsiveness, while offline label metrics also improve modestly (Sec.~\ref{sec:ablation}).

\section{Experiments}\label{sec:experiments}

\paragraph{Scope of evaluation.} Given the largely teacher-generated targets described in Sec.~\ref{sec:data_synthesis}, the offline metrics in Sec.~\ref{sec:main_results} measure agreement with the dataset targets and the structural validity of generated sequences. The LLM judge in Sec.~\ref{sec:indep_eval} evaluates semantic plan quality from the user context without seeing those targets, while the online A/B test in Sec.~\ref{sec:online} measures changes in UV-CTR relative to the incumbent ranker.

\subsection{Experimental Settings}

\subsubsection{Datasets and Baselines.}
We conduct experiments on a large-scale industrial dataset comprising 100{,}000 training intent sequences and 1{,}000 test sequences from Amap, constructed via the pipeline described in Sec.~\ref{sec:data_synthesis}. We compare GPlan (1.7B and 4B versions) against two categories of baselines:
(1) traditional sequential models: SASRec \cite{SASRec}, BERT4Rec \cite{BERT4Rec} and a reasoning-based model ReaRec \cite{ReaRec};
(2) generative recommendation models: P5 \cite{P5}, TIGER \cite{TIGER}. We adapt each baseline to the full intent sequence task using its original architecture and inputs: all baselines use historical intent sequences, while P5 also uses user profiles and spatiotemporal context. Since the baselines do not generate parameter slots, Table~\ref{tab:main_results} evaluates all methods at the tool level. Our GPlan models use the Qwen3 1.7B and 4B backbones.

\subsubsection{Evaluation Metrics}\label{sec:eval_metrics}
We evaluate GSISR offline via three dimensions: (1) Acc@1, measuring the accuracy of the first intent, which is prioritized in Amap's homepage UI; (2) NDCG \cite{10.1145/345508.345545}, assessing the ranking relevance of the entire sequence; and (3) normalized edit similarity (NES), which evaluates the structural alignment between the predicted and ground-truth intent sequences. It is defined as:
\begin{equation}
\text{NES}(\hat{\mathbf{Y}}, \mathbf{Y}^*) = 1 - \frac{D(\hat{\mathbf{Y}}, \mathbf{Y}^*)}{\max(|\hat{\mathbf{Y}}|, |\mathbf{Y}^*|)}
\end{equation}
where $D$ is the weighted Levenshtein edit distance with insertion and deletion costs of $1.0$, a tool-level substitution cost of $1.0$ (different tool name), and a parameter-level substitution cost of $0.3$ (same tool, different parameters). $\text{NES} \in [0,1]$, where higher values indicate better sequence-level alignment.

\subsubsection{Implementation Details.} GPlan uses the Qwen3 1.7B and 4B backbones. PICD trains for $13$ epochs with $K=3$ latent tokens per block, $B^\star = 9$, $\eta_{\text{struct}} = 5\!\times\!10^{-6}$, and $\eta_{\text{polish}} = 1\!\times\!10^{-6}$. SC-DPO is initialized from the PICD-CALR checkpoint and trained for $1$ epoch on the counterfactual pair corpus at learning rate $2\!\times\!10^{-7}$ (constant with warmup ratio $0.03$), with $\beta_{\text{DPO}} = 0.20$, anchor margin $\delta = 0$ and weight $\lambda_a = 5$, gap range $[\gamma_\ell, \gamma_h] = [0.10, 0.20]$ with asymmetric weights $\lambda_{g,\ell} = 10$ on the lower bound and $\lambda_{g,h} = 2$ on the upper bound, and center target $m = 0$ with weight $\lambda_c = 3$.

\paragraph{Latent-Structure Validity.}\label{sec:latent_diag} In addition to task metrics, we report a deterministic diagnostic specific to PICD that measures whether the emitted latent reasoning prefix is well-formed. A predicted output is counted as latent-valid if it contains a single
\texttt{<THOUGHT>...</THOUGHT>} block whose latent tokens (i) follow the scaffold order $C,S,T_1,\ldots,T_{|\mathbf{Y}|}$ defined in Sec.~\ref{sec:picd_scaffold}, (ii) contain no repeated or out-of-order tokens, (iii) leave no residual raw XML tags such as \texttt{<STEP\_$i$>}, and (iv) yield a step count equal to the JSON plan length $|\mathbf{Y}|$. We use this diagnostic to surface latent-structure failures that JSON-based metrics cannot detect (Sec.~\ref{sec:abl_calr}).

\subsection{Main Results}\label{sec:main_results}
As shown in Table~\ref{tab:main_results}, GPlan at both model sizes consistently outperforms the traditional sequential and generative recommendation baselines. The gains in both Acc@1 and NDCG@3, including GPlan-4B's $18.5$ percentage-point margin over TIGER in Acc@1, suggest that existing recommenders still struggle to capture the complex spatiotemporal constraints and logical dependencies of the full intent sequence task. Notably, GPlan-1.7B retains the bulk of GPlan-4B's accuracy ($60.5$ vs.\ $62.7$ Acc@1), demonstrating that our distillation strategies effectively internalize complex reasoning capabilities into smaller architectures.

\begin{table}[!t]
\centering
\caption{Performance Comparison on Industrial Dataset.}
\label{tab:main_results}
\begin{tabular}{l|c|c}
\hline
\textbf{Method} & \textbf{Acc@1 ($\uparrow$)} & \textbf{NDCG@3 ($\uparrow$)}\\ \hline
SASRec & 36.1 & 0.5865  \\
BERT4Rec & 37.4 & 0.5973  \\
ReaRec & 42.9 & 0.6442  \\
P5 & 43.5 & 0.6581  \\
TIGER & 44.2 & 0.6647  \\
\hline
\textbf{GPlan-1.7B} & 60.5 & 0.7450  \\
\textbf{GPlan-4B} & \textbf{62.7} & \textbf{0.7759} \\ \hline
\end{tabular}

\end{table}

\subsection{Independent Evaluation by LLM Judge}\label{sec:indep_eval}
Label-based metrics measure how closely a model reproduces curated targets, but not whether the resulting plan is itself a good recommendation. We therefore complement them with an LLM judge (GPT-4o, temperature 0) that scores each candidate sequence on three semantic axes directly tied to GPlan's claims. The judge sees only the user context and the candidate plan, never the curated target sequence. All GPlan variants additionally pass the rule-based filter of Sec.~\ref{sec:data_synthesis} (JSON, schema, mutual-exclusion, no-repetition), so we focus the LLM-judge tables on the more discriminating semantic axes.

The rubric is calibrated so that a safe but generic plan receives \textbf{3}, not 5: getting above 3 requires positive evidence on the axis, while a score below 3 requires an explicit failure. This quality-driven calibration produces a meaningful spread of scores across models rather than ceiling saturation, which we observed under earlier failure-driven rubrics that defaulted to 5.

\aptLtoX{\par\smallskip
\noindent
\begingroup
\small
\setlength{\tabcolsep}{6pt}
\renewcommand{\arraystretch}{1.1}

\begin{tabular}{|>{\raggedright\arraybackslash}l{\dimexpr\columnwidth-14pt\relax}|}
  \hline
  \rowcolor{gray!65!black}\color{white}\bfseries 
  Judge Rubric Skeleton \\
  \hline
  \rowcolor{gray!4!white}
  \textbf{Three axes (scored independently, 1--5):}\\
  \begin{itemize}\itemsep1pt
  \item \emph{Flow.} Does the first intent hit the user's most immediate need, and do later intents extend that anchor with personalized signals? \emph{Positive: first intent matches the immediate anchor + personalized continuation. Failure: missed anchor or topic drift.}
  \item \emph{Logic.} Does the ordering reflect a service-lifecycle progression (arrival\,$\to$\,on-site\,$\to$\,post-service) rather than an arbitrary permutation? \emph{Positive: clean stage progression (urgent/certain\,$\to$\,arrival\,$\to$\,on-site\,$\to$\,post-service). Failure: stage inversion or dependency error.}
  \item \emph{Spatiotemporal.} Do tools, parameters, and tag/range choices specifically fit the current time, POI, and order state, or would they look the same in any context? \emph{Positive: parameters/tools accurately fit current time, POI, and order state. Failure: parameter mismatch or ignored order/spatial state.}
  \end{itemize} \\
    \rowcolor{gray!4!white}
  \textbf{Five-level anchors (applied per axis, using the per-axis cues above):}\\
  \begin{itemize}\itemsep1pt
  \item \textbf{5} positive evidence on this axis
  \item \textbf{4} sound on this axis, with one clear positive signal
  \item \textbf{3} no failure but generic / template-like (the default)
  \item \textbf{2} one explicit failure on this axis
  \item \textbf{1} systemic failure across the axis or no shared goal
  \end{itemize} \\
  \hline
\end{tabular}
\endgroup
\par\smallskip}{
\par\smallskip
\begingroup
\setlength{\fboxsep}{6pt}
\setlength{\fboxrule}{0.5pt}
\begin{lrbox}{\gplanpanelbody}
\begin{minipage}{\dimexpr\columnwidth-2\fboxsep-2\fboxrule-12pt\relax}
\small
\textbf{Three axes (scored independently, 1--5):}
\begin{itemize}\itemsep1pt
\item \emph{Flow.} Does the first intent hit the user's most immediate need, and do later intents extend that anchor with personalized signals? \emph{Positive: first intent matches the immediate anchor + personalized continuation. Failure: missed anchor or topic drift.}
\item \emph{Logic.} Does the ordering reflect a service-lifecycle progression (arrival\,$\to$\,on-site\,$\to$\,post-service) rather than an arbitrary permutation? \emph{Positive: clean stage progression (urgent/certain\,$\to$\,arrival\,$\to$\,on-site\,$\to$\,post-service). Failure: stage inversion or dependency error.}
\item \emph{Spatiotemporal.} Do tools, parameters, and tag/range choices specifically fit the current time, POI, and order state, or would they look the same in any context? \emph{Positive: parameters/tools accurately fit current time, POI, and order state. Failure: parameter mismatch or ignored order/spatial state.}
\end{itemize}

\textbf{Five-level anchors (applied per axis, using the per-axis cues above):}
\begin{itemize}\itemsep1pt
\item \textbf{5} positive evidence on this axis
\item \textbf{4} sound on this axis, with one clear positive signal
\item \textbf{3} no failure but generic / template-like (the default)
\item \textbf{2} one explicit failure on this axis
\item \textbf{1} systemic failure across the axis or no shared goal
\end{itemize}
\end{minipage}
\end{lrbox}

\noindent
\begin{minipage}{\columnwidth}
\fcolorbox{gray!65!black}{gray!65!black}{%
  \makebox[\dimexpr\columnwidth-2\fboxsep-2\fboxrule\relax][l]{%
    \hspace*{6pt}\color{white}\bfseries\small
    Judge Rubric Skeleton}}%
\par\nointerlineskip
\kern-\fboxrule
\noindent
\fcolorbox{gray!65!black}{gray!4!white}{%
  \makebox[\dimexpr\columnwidth-2\fboxsep-2\fboxrule\relax][c]{%
    \usebox{\gplanpanelbody}}}%
\end{minipage}
\endgroup
\par\smallskip}

A residual risk is that an LLM judge may reward superficial mentions of context (e.g., a plan that names the current POI without actually using it for tool or parameter selection) even when the plan does not truly use the context. We reduce, but do not eliminate, this risk by scoring concrete axes that inspect ordering, tool choice, parameter ranges, and fit to the current POI/time/order state. The online A/B result in Sec.~\ref{sec:online} provides a complementary signal that does not rely on LLM judgment.

Because all baselines emit ID-only sequences without the parameter slots the rubric inspects, the judge is applied only within the GPlan family, where it differentiates training paradigms rather than backbone capacity. The per-axis scores together with their mean \emph{Overall} are reported alongside the label-based metrics in Table~\ref{tab:ablation_strategy} (Sec.~\ref{sec:ablation}); a targeted slice for the SC-DPO design appears in Table~\ref{tab:ablation_scdpo}.

\subsection{Ablation Analysis}\label{sec:ablation}

\paragraph{Effectiveness of Learning Strategies.}
Using the same Qwen3-4B backbone, we compare Direct-SFT, CoT-SFT, PICD, and PICD with SC-DPO. Table~\ref{tab:ablation_strategy} reports the label-based metrics together with the Flow, Logic, Spatiotemporal (ST), and Overall LLM-judge scores. PICD is close to CoT-SFT on the label-based metrics and is slightly higher on all three judge axes. Adding SC-DPO mainly increases the ST score from $3.69$ to $3.75$, while the remaining differences are small.

\begin{table*}[!t]
\centering
\caption{Ablation of post-training strategies on GPlan-4B.}
\label{tab:ablation_strategy}
\begin{tabular}{l|c|c|c|c|c|c|c}
\hline
\multirow{2}{*}{\textbf{Method}} & \multicolumn{3}{c|}{\textbf{Label-based metrics}} & \multicolumn{4}{c}{\textbf{LLM judge}} \\
\cline{2-8}
 & \textbf{Acc@1} & \textbf{NDCG@3} & \textbf{NES} & \textbf{Flow} & \textbf{Logic} & \textbf{ST} & \textbf{Overall} \\ \hline
Direct-SFT             & 61.2 & 0.7591 & 0.6210 & 3.37 & 3.43 & 3.49 & 3.43 \\
CoT-SFT                & 62.3 & 0.7785 & 0.6354 & 3.53 & 3.61 & 3.64 & 3.60 \\
\textbf{PICD}          & 62.6 & 0.7726 & 0.6290 & 3.56 & 3.62 & 3.69 & 3.62 \\
\textbf{PICD + SC-DPO} & 62.7 & 0.7759 & 0.6365 & 3.58 & 3.59 & 3.75 & 3.64 \\ \hline
\end{tabular}
\end{table*}

\paragraph{Compression-Aware LR vs.\ Single-Phase Schedules.}\label{sec:abl_calr}
Table~\ref{tab:ablation_calr} compares three learning-rate schedules for PICD (Sec.~\ref{sec:picd_calr}). Cosine LR follows the standard monotonic-decay schedule and lowers the learning rate before the latent scaffold has been fully acquired. Constant LR keeps the structure-learning rate throughout training, preserving the scaffold but leaving no lower-LR phase for refining the JSON plan. CALR explicitly separates scaffold acquisition from plan refinement. Alongside the sequence-level label metrics, \emph{Latent-Valid} (Sec.~\ref{sec:latent_diag}) indicates whether the implicit-CoT interface is preserved end to end. Cosine LR achieves competitive task metrics, but its emitted prefix no longer satisfies the scaffold: the model recovers a serviceable JSON plan while bypassing the intended latent interface. Constant LR and CALR both preserve the scaffold; among them, CALR also achieves the strongest label-based results by retaining a lower-LR phase for plan refinement.

\begin{table}[!htbp]
\centering
\caption{PICD learning-rate ablation on GPlan-4B.}
\label{tab:ablation_calr}
\begin{tabular}{l|c|c|c|c}
\hline
\textbf{Schedule} & \textbf{Acc@1} & \textbf{NDCG@3} & \textbf{NES} & \textbf{Latent-Valid} \\ \hline
Cosine-LR & 62.4 & 0.7652 & 0.6259 & 0.0\\
Constant-LR & 62.1 & 0.7575 & 0.6216 & 1.0\\
\textbf{CALR} & 62.6 & 0.7726 & 0.6290 & 1.0 \\ \hline
\end{tabular}
\end{table}

\paragraph{Reference-Anchored vs.\ Vanilla DPO.}\label{sec:abl_scdpo}
To isolate the contribution of the reference-anchored design in SC-DPO (Sec.~\ref{sec:scdpo}), we ablate against (i) no DPO at all (PICD with CALR) and (ii) a \emph{vanilla DPO} variant that omits the anchor / gap / center regularization terms but uses the same counterfactual pair construction. Both DPO variants start from the same PICD-with-CALR checkpoint and use the same counterfactual training corpus; only their post-training losses differ. Beyond label-based metrics, we report the LLM-judge \textbf{ST} axis --- the axis most directly tied to SC-DPO's grounding motivation --- and the judge \emph{Overall} to check whether improvements on ST come at the cost of the other two axes.

\begin{table}[!htbp]
\centering
\caption{SC-DPO loss-design ablation on GPlan-4B.}
\label{tab:ablation_scdpo}
\begin{tabular}{l|c|c|c|c|c}
\hline
\textbf{Post-training} & \textbf{Acc@1} & \textbf{NDCG@3} & \textbf{NES} & \textbf{ST} & \textbf{Overall} \\ \hline
PICD & 62.6 & 0.7726 & 0.6290 & 3.69 & 3.62 \\
+ Vanilla DPO           & 62.1 & 0.7484 & 0.6158 & 3.60 & 3.51 \\
+ \textbf{SC-DPO} & 62.7 & 0.7759 & 0.6365 & 3.75 & 3.64 \\ \hline
\end{tabular}
\end{table}

Table~\ref{tab:ablation_scdpo} isolates whether the combined anchor / gap / center regularization preserves the SFT-acquired matched-response quality while introducing context-conditional preference. SC-DPO treats $y_-$ as \emph{context-mismatched} rather than globally low-quality, so its primary operational signature is a stronger judge ST score without reducing the judge Overall or the label-based metrics. Vanilla DPO, by pushing $\pi_\theta(y_- \mid x)$ down without constraint, instead drifts downward across all reported axes, confirming the need for this combined regularization.

\paragraph{Inference Efficiency.} We evaluate the deployment feasibility by measuring mean response time (RT) across the deployable scales and methods. All numbers are measured on two NVIDIA H20 GPUs under a target serving load of QPS $=10$, with greedy decoding. As shown in Table~\ref{tab:rt_comp}, while CoT-SFT provides strong reasoning, its latency (up to $1.85$\,s at 4B) is incompatible with real-time production. Our PICD achieves a $3.7\times$ speedup over CoT-SFT at the 4B scale, maintaining an RT close to Direct-SFT. This gives PICD a favorable accuracy/latency operating point for industrial deployment.

\begin{table}[!htbp]
\centering
\caption{Inference latency (mean RT) across model scales and post-training methods.}
\label{tab:rt_comp}
\begin{tabular}{l|c|c|c}
\hline
\textbf{Scale} & \textbf{Direct-SFT} & \textbf{CoT-SFT} & \textbf{PICD (Ours)} \\ \hline
1.7B & 299\,ms & 1.06\,s & 366\,ms \\
4B   & 437\,ms & 1.85\,s & 503\,ms \\ \hline
\end{tabular}
\end{table}

\subsection{Online A/B Testing}\label{sec:online}
GPlan was deployed to the Amap homepage in an A/B test against the incumbent CTR-based ranking model, using GPlan-1.7B as the serving model. The test ran for 14 consecutive days. Two parallel slices were run: a general slice on $5\%$ of homepage traffic ($\sim 600$ QPS), and a separate cross-city slice on another $5\%$ that isolates users with an active out-of-town context. The general slice yielded a $+0.87\%$ UV-CTR lift over the production baseline, while the cross-city slice yielded a $+1.04\%$ UV-CTR lift, indicating that GPlan's gains are larger in high-value spatiotemporal scenarios where sequence-level planning matters most.

\section{Conclusion}
We formalized \textbf{GSISR} --- generating coherent, physically executable intent sequences under spatiotemporal constraints --- and proposed \textbf{GPlan} to internalize LLM-level planning into deployable students. PICD combines a multi-token latent reasoning scaffold with a \emph{compression-aware learning rate} that ties LR transitions to the state of the latent compression curriculum, preserving the implicit-CoT interface while refining the JSON plan at production latency. SC-DPO introduces reference-anchor, bounded-margin, and centered-drift terms on top of counterfactual preference pairs, so the model learns context-conditional preference without degrading the SFT-acquired planning ability. Online A/B tests on Amap homepage confirm that GPlan delivers measurable user-side gains over the production CTR-based ranker, with a larger lift on high-value cross-city scenarios where sequence-level planning matters most. Future work will explore reinforcement learning and multi-turn user feedback to further improve personalization and long-horizon adaptability.

\bibliographystyle{ACM-Reference-Format}
\bibliography{sample-base}

\end{document}